\documentstyle[preprint,prl,aps]{revtex}          

\begin{document}

\title{Absence of backscattering at integrable impurities 
in one-dimensional quantum many-body systems}

\author{Alexander Punnoose}
\address{Department of Physics,
Indian Institute of Science,
Bangalore, 560 012, India}

\author{Hans-Peter Eckle}
\address{LMPM,
D\'{e}partement de Physique, 
Universit\'{e} Fran\c{c}ois Rabelais,
F-37200 Tours, France}

\author{Rudolf A.\ R\"{o}mer\cite{rar}}
\address{Institut f\"ur Theoretische Physik C, 
RWTH Aachen,
Templergraben 55, 
D-52056 Aachen, Germany}

\date{Version: December 18, 1995; printed \today}
\maketitle

\begin{abstract}
We study interacting one dimensional (1D) quantum lattice gases
with integrable impurities. These model Hamiltonians can be
derived using the quantum inverse scattering method for
inhomogeneous models and are by construction integrable. 
Absence of backscattering at the impurities is shown to be the 
characteristic feature of these disordered systems. 
The value of the effective carrier charge and the
Sutherland-Shastry relation are derived for the half-filled
XXX model and are shown to be independent of the
impurity concentration and strength. 
For the half-filled XXZ model
we show that there is no enhancement of the persistent currents 
for repulsive interactions. For attractive interactions
we identify a crossover regime beyond which enhancement of the
currents is observed. 
\end{abstract}


%
%


Much work has been done recently towards a deeper understanding of 
the effect of impurities in Fermi- and Luttinger liquids.
Conformal field theories with boundaries have been used to 
study the multichannel Kondo problem \cite{affleck}. 
Bosonization techniques and renormalization group methods have
given new insights into the problem of potential scatterers in
Luttinger liquids \cite{tgh1,tgh2,kane}.
Also, the observed discrepancy between the experimental value of the 
persistent current
in moderately disordered ensembles of quasi one-dimensional (1D) rings
\cite{levy} and the theoretical predictions
based on calculations in a disordered but non-interacting electron gas
\cite{boris}, have generated much interest in the interplay
between interactions and impurities
\cite{rarap,tgbss,ramin,assa,mori}. 

Besides the spin impurity model of Andrei and Johannesson
\cite{ja}, no other {\em exactly solvable} interacting
quantum many-body problems {\em with} impurities have been
studied. This is in retrospect somewhat surprising as 
the existence of models with site-dependent defects has been mentioned
throughout the literature on the method of quantum inverse
scattering (QISM) \cite{baxter,taktajan,shastry,sklyanin,korepin}.
Recently, Bares \cite{bares} and, independently, 
Schmitteckert, Schwab and Eckern (SSE) \cite{eckern} 
have used these remarks and explicitly investigated a
$tJ$ model and an XXZ model with integrable defects,
respectively. 

In the XXZ model, SSE find the following surprising properties: 
(i) The energy spectrum is independent of the spatial distribution of 
impurities; 
(ii) there is no localisation of the ground state wave function.
In contrast measurements of the transport properties in mesoscopic
systems show a very sensitive dependence on the impurity
configuration \cite{exp}. 
Further, it is well understood that in the thermodynamic limit in 1D
a generic impurity drives the system from a metallic to an insulating
phase even in the presence of weak interactions
\cite{tgh1,tgh2,a1}.  

In the present work, we shall first introduce 
the (antiferromagnetic) XXX Heisenberg model as a simple example of a system
with integrable impurities.
We will then show that a complete physical understanding of the
anomalous properties of this class of models can be got by
studying the properties of a single particle scattering off
a single impurity.  The system is then threaded with an
external flux $\Phi$.  
For the half-filled band, we establish the charge of the
effective carriers to be $-q/2$. The Sutherland-Shastry
relation \cite{bs,bss} $D \chi = 1/2\pi$ still holds in the presence of
impurities. 
For the XXZ model with integrable impurities at half
filling the charge
stiffness $D$ shows
enhancement of the persistent currents for 
attractive interactions in agreement with studies of XXZ
models with potential impurities \cite{ramin}. 

%
%


We construct our model from the $R$ and $L_n$ matrices of the
Heisenberg XXX model \cite{taktajan}, i.e.\
$R(\lambda)= (\lambda P + i)/(\lambda+i)$ and
$L_n(\lambda)= \lambda + \frac{i}{2} 
 \sum_{a=1}^{3} \sigma^a \otimes \sigma_n^a$.
$P$ is a $4\times 4$ permutation matrix and $\{\sigma^a\}$ are the Pauli
matrices. These operators satisfy the local Yang-Baxter (YB)
equation, i.e.\ 
$R(\lambda-\lambda ') [L_n(\lambda)\otimes L_n(\lambda')]= 
[L_n(\lambda')\otimes L_n(\lambda)]R(\lambda-\lambda')$.
The main idea that allows for the construction of models with
integrable impurities is that
the YB equation continues to be satisfied under an arbitrary {\em
local} shift in the spectral parameter 
$L_n(\lambda) \rightarrow L_n(\lambda+\nu_n)$.
The transfer matrix $T$ with local shifts 
$\{\nu_1,\cdots,\nu_N\}$ is
given as 
$T(\lambda, \{\nu_n\}) = \prod_{n=1}^{N}L_n(\lambda+\nu_n)$.
The Hamiltonian $H$ is then given as usual by
the logarithmic derivative of the transfer matrix $
T(\lambda,\{\nu_n\}) $ evaluated at the special point $\lambda_0 =i/2$.
Note that this construction is different from the one used by
Andrei and Johannesson \cite{ja} and does indeed yield inequivalent,
albeit similar, Bethe Ansatz (BA) equations.

As is customary, we now Jordan-Wigner transform the spin model 
into a lattice gas of spinless fermions and impose periodic
boundary conditions.
The resulting Hamiltonian on a ring of size $N$
with $M$ spinless fermions and a single impurity at site $m$
with strength $\nu$
is given as $H = H_{XXX} + H_{m,imp}$, where 
$H_{XXX}= 
- \sum_{j=1}^{N} (c_{j+1}^{\dagger}c_{j}+ h.c.) + 
   2 \sum_{j=1}^{N} \rho_{j} \rho_{j+1}  + 2M
$
is the usual Hamiltonian of the clean XXX model in units of the
hopping energy.
The operators $c_{j}^{\dagger}$ and $c_{j}$ create and
annihilate fermions on 
site $j$ and $\rho_{j} = c_{j}^{\dagger}c_{j}$ is the density
operator.
The impurity interaction at site $m$ is 
$H_{m,imp} = H_{m,t} + H_{m,\rho} + H_{m,j}$ with 
\begin{mathletters}
\label{eqn-himp}
\begin{eqnarray}
H_{m,t} &= &-(u + i w) 
   [ c_m^{\dagger}c_{m-1}+c_{m+1}^{\dagger}c_m+
   c_{m-1}^{\dagger}c_{m+1} ] \nonumber \\ && \mbox{ }
   + h.c.,
\label{eqn-ht} \\
H_{m,\rho} &= &
    -2 u [ \rho_{m} - \rho_{m}(\rho_{m-1}+\rho_{m+1}) + 
     \rho_{m-1}\rho_{m+1} ],
\label{eqn-hrho} \\
H_{m,j} &=&
     2 u
     ( c_{m+1}^{\dagger}c_{m-1} + c_{m-1}^{\dagger}c_{m+1} ) 
     \rho_{m}  \nonumber \\ && \mbox{ }
   + 2iw 
    ( c_{m}^{\dagger}c_{m-1} - c_{m-1}^{\dagger}c_{m} )
    \rho_{m+1} \nonumber \\ && \mbox{ } 
   + 2iw 
    ( c_{m+1}^{\dagger}c_{m} - c_{m}^{\dagger}c_{m+1} )
    \rho_{m-1},
\label{eqn-hj}
\end{eqnarray}
\end{mathletters} 
where the coupling constants as a function of $\nu$ are
$u= -\nu^2/(1+\nu^2)$ and 
$w= -\nu/(1+\nu^2)$.
In Eq.\ (\ref{eqn-himp}), we have three types of terms:
(i) $H_{m,t}$: Nearest and next nearest neighbour hopping matrix
elements are  modified involving the sites $m$ and $m\pm 1$.
(ii) $H_{m,\rho}$: The onsite potential at site $m$ and both 
nearest and next nearest neighbour density-density interaction terms
are induced. 
(iii) $H_{m,j}$: This term corresponds to a current density
interaction.
Thus we see that a local translation of the spectral parameter 
$\lambda\rightarrow \lambda+\nu_m$ effectively involves the site
$m$ and its two neighbors $m\pm 1$ and in addition involves next
nearest neighbour terms involving the sites $m-1$ and $m+1$.
We note that the impurity terms involve qualitatively
the same processes as in the $tJ$ model with integrable impurity
\cite{bares}.

The Hamiltonian in the presence of more than one impurity can now
easily be constructed: In case the impurities are well separated 
($|m-m'|>1$)
the Hamiltonian reduces to a
sum over the isolated impurities
i.e.\ $\sum_{\alpha=1}^{F} H_{m(\alpha),imp}$ for
$F$ such well-separated impurities at positions 
$m(\alpha)$, $\alpha=1, \ldots, F$.
In case two impurities are on neighboring sites, say $m$ and $m+1$,
the Hamiltonian will acquire similar
terms involving the sites $m-1$ to $m+2$. As we will show later,
this does not change any of the physics of these systems.

%
%


To understand the properties of integrable impurities, we first
study the scattering of a single particle by a single impurity at
site $m$. 
The terms in the Hamiltonian involving only a single
particle then are 
$H_s=-\sum_{i}(c_i^{\dagger}c_{i+1}+h.c.) + H_{t} -2 u \rho_m + 2$.
Let $\Psi_{in}(n)= e^{i k n} + R(k) e^{-i k n}$ for $ n< m $ be the
incoming and $\Psi_{out}(n)= T(k) e^{i k n}$ for $n > m$ be the
outgoing wave functions and 
$R(k)$ and $T(k)$ the reflection and transmission amplitudes,
respectively. Then with $\epsilon(k)= -2(\cos k + 1)$ the usual
nearest-neighbor one-particle dispersion, we find that
(i) $R(k)=0$ for all momenta $k$.
Thus there is {\em no backscattering} of a particle due to the impurity.
This is consistent with the understanding that the number of scattering 
channels in a BA-solvable Hamiltonian is conserved. 
In contrast, a generic impurity would give rise to a reflected and a 
transmitted wave thus doubling the number of scattering channels.
We remark that the parametrisation of $u$ and $w$ given in
Eq.\ (\ref{eqn-himp}) is unique, i.e.\ it is the only such parametrisation 
that will lead to $R(k)=0$ for all $k$ and $\nu$.
(ii) Unitarity of the scattering matrix requires $|T|^2+|R|^2=1$ and we
consequently have $|T(k)|^2=1$. Writing $T(k)= e^{i \delta(k,\nu)}$,
we then find the phase-shift of the particle wave function due to the
presence of the impurity to be
$\delta(k,\nu)= -k + 2 \arctan(\tan(k/2)-2\nu)$.

To study the problem with more than one particle, we use the BA
equations in the presence of the impurities which are given for the
XXX model by QISM as \cite{taktajan} 
\begin{equation}
N k_j=  2\pi I_{j} -
\sum_{k=1}^{M}\theta(k_{j},k_{k})
-  \sum_{\alpha=1}^{F} \delta(k_j,\nu_{m(\alpha)}), 
\label{eqn-baXXXimp}
\end{equation} 
where $\theta(k)= 2\arctan[ (\tan(k/2)-\tan(k'/2) )/2 ]$ is the
two-body phase shift.
Note that although the total momentum $P$
is no longer a good 
quantum number in the presence of impurities, the individual
BA pseudo-momenta $k$ still represent the eigenvalues of the
BA equations and are thus good quantum numbers.

We now make the important observation that 
the only change in the BA equations w.r.t.\ the
clean case is the additional phase shift $\delta(k,\nu)$ which is
{\em identical} to the forward scattering phase shift acquired by a
single particle scattering off the impurity. This is true although
the BA equations include many-body interactions and the interactions 
$H_{m(\alpha),\rho}$ and $H_{m(\alpha),j}$ induced by the impurities.
Thus it is also irrelevant, if the impurities are well separated or
not. For all configurations of the impurity the net effect
will be the same phase shift $\delta(k,\nu_m)$ for each impurity of
strength $\nu_m$. 
Further, Eq.\ (\ref{eqn-baXXXimp}) is insensitive to the spatial
distribution of the $F$ impurities. This is a
consequence of the absence of backscattering. Quantum
interference of multiply backscattered waves would in general
lead to a non trivial dependence of the energy spectrum on the
impurity configuration.
Note that, although the energy spectrum is insensitive to the
impurity configuration the wave function 
is no longer given by the simple coordinate BA
form $\Psi(x_1, \ldots, x_M)= \sum_{P} A(P) \exp( i \sum_{j=1}^{M}
x_j k_{Pj} )$ with $P$ a permutation of the $M$ momenta ${k_j}$.   
Presumably, one might calculate the wave function from the reference
state of the QISM, which then by construction includes a site 
dependence.

We believe that these results are generic for all such integrable
impurities constructed by QISM. This is implied by the very existence
of the BA equations.
Besides the XXX model, the $tJ$ model of Bares \cite{bares}
and the XXZ ring of SSE \cite{note1}, this should therefore also
apply to, e.g., the Hubbard model with integrable impurities constructed
from two coupled inhomogeneous six-vertex models \cite{shastry}.

%
%


To study the transport properties of these integrable impurity
models, we thread the system with an external flux $\Phi$, which 
accelerates all the particles around the ring.
The variation of the energy spectrum as a function of flux
is made nontrivial for an interacting quantum system only in the
absence of Galilean invariance.
In the present problem this invariance is destroyed both due to the
presence of the lattice {\em and} the existence of the impurities.

The ground state energy $E_0(\Phi)$ for a given flux $\Phi$ is 
periodic in $\Phi$ with period $2\pi$. However, the adiabatic
variation of the energy $E(\Phi)$ of 
the state that begins as the ground state at $\Phi=0$ may have a
periodicity greater than $2\pi$.  
The clean ($\nu=0$) XXX model has been studied 
previously by Sutherland and Shastry \cite{bss}. 
They find a periodicity of $4\pi$
implying carriers with charge $-q/2$ in terms of the
charge $q$ of the fundamental carriers. 
The second quantity of interest is the stiffness 
$D=(N/2)\partial^2{E_0(\Phi)}/\partial\Phi^2|_{\Phi=0}$. Disorder
will in general destroy the phase sensitivity of the ground
state and therefore lead to a non trivial renormalization of the
stiffness.

Let us consider the half-filled case $N=2M$.
In order to study the periodicity of the ground state,
we follow the particle with the maximum momentum $k_M$
characterised by the ground state quantum number $I_M=(M-1)/2$
as a function of $\Phi$.
At the special point $k_M=\pi$, the BA Eq.\ (\ref{eqn-baXXXimp})
for $k_M$ with $\theta(k_i,\pi)=\pi$ for 
$i<M$ and $\delta(\pi,\nu)=0$ for all finite $\nu$ gives the
flux to be $2\pi$.  Conversely, we notice that as long as
$|\Phi| < 2\pi $, all $k'\mbox{s}$ stay within $|k|<\pi$ until
at $\Phi=2\pi$, the maximum momentum takes the value $k_M=\pi$.
Hence, it follows that for $2\pi<\Phi<4\pi$ the $k'\mbox{s}$ are
the same as in the case $|\Phi|<2\pi$ with the states relabelled
as $k_{M-1}>k_{M-2}>\cdots>k_1>k_M$ modulo $2\pi$.
We next substitute the values $k_M=\pi$ and $\Phi=2\pi$ into the
remaining $M-1$ equations. These reduce to 
\begin{equation}
N k_{j} = 2\pi I_{j}' - \sum_{l\neq M}\theta(k_{j},k_{l})
 -  \sum_{\alpha=1}^{F}\delta(k_j,\nu_{m(\alpha)}).
\label{eqn-dchi}
\end{equation} 
We immediately notice that these equations with the quantum
numbers $I_{j}'=I_{j}+\frac{1}{2}$ are just the BA equations
for the ground state of $M-1$ particles in the absence of flux.
The energy of the ground state of the $M$ particle system 
$E(\Phi)$ at flux $\Phi=2\pi$  is thus related to the
ground state of $M-1$ particles at $\Phi=0$, i.e.\
$E_{M}(2\pi)= E_{0,M-1}(0)-2(1+\cos(k_M))=E_{0,M-1}(0)$.
Therefore
\begin{eqnarray}
\Delta E_{M}(\Phi)
&=& E_{M}(2\pi) - E_{M}(0) = 4 \pi^2 D/N \nonumber \\  
&=& E_{0,M-1}(0)-E_{0,M}(0) = 2 \pi \chi^{-1}/N,
\end{eqnarray}
with $\chi$ the susceptibility. This remarkable property of the
BA Eq.\ ({\ref{eqn-baXXXimp}) relates the stiffness and the
susceptibility through $D \chi = 1/2\pi$
even in the presence of impurities.
This is the Sutherland-Shastry relation derived previously in the
clean XXZ \cite{bs},
Hubbard \cite{bss} and SC model \cite{rarbs}.

We next turn to explicitly calculating the stiffness $D$ of a
system with integrable impurities. For convenience, we will use
the XXZ spinless fermion model with impurities where we can
easily vary the interaction strength $\Delta=\cos(2\eta)$. The
clean system is given by the Hamiltonian, 
$
H_{XXZ}= 
- \sum_{j=1}^{N} (c_{j+1}^{\dagger}c_{j}+ h.c.) -
  2 \Delta \sum_{j=1}^{N} \rho_{j} \rho_{j+1}.
$
The stiffness is calculated by studying the 
leading order finite size corrections to the thermodynamic
value of the ground state energy in the presence of a flux
with the help of the Wiener-Hopf technique \cite{peter}. 
For the XXZ model, this was already done by SSE for impurities
chosen symmetric w.r.t.\ the origin and of equal strength $\nu$. 
They get for the  stiffness 
\begin{equation}
D(\eta,\nu)=
 \frac{\pi}{8}
 \frac{\sin(2\eta)/(\eta(\pi-2\eta))}{1-f+f\cosh[\pi\nu/(\pi-2\eta)]}.
\label{eqn-stiffness}
\end{equation}
where $f= F/N$ is the impurity density. 
By the same arguments as above, $D\chi=1/2\pi$ also in the XXZ model
and we have the susceptibility in the presence of impurities.
Using hydrodynamical arguments, we further have 
$D \chi^{-1} = v^2$ \cite{billth}, where $v$ is the Fermi velocity.
Thus we see that the impurities simply renormalize the values
of the stiffness, inverse susceptibility and the Fermi
velocity of the clean system by the factor
$(1-f+f\cosh[\pi\nu/(\pi-2\eta)])^{-1}$. 
This then further implies that the critical exponents remain unchanged
and that the correlation functions are modified by the 
renormalization of $v$ only.
We remark here that these calculations can also be done by choosing a 
more general, say Gaussian, distribution for the impurity strengths.
However, the resulting expressions are lengthy and will not be included in
this letter.

A finite stiffness also implies that the ground state is extended. 
The stiffness in the $\nu\neq 0$ case as seen from
Eq.\ (\ref{eqn-stiffness}) has a quadratic dependence on the
strength of the impurity and only in the limit of very large
disorder does one get an exponential dependence. 
This is contrary to the understanding that due to coherent back
scattering in 1D a generic
impurity would lead to localisation for arbitrarily weak
disorder.
However this result is consistent with the observation that
these are "special" impurities with zero reflection. 

The stiffness provides an operational definition for the
persistent currents as $J= D \Phi$.
In Fig.\ \ref{fig}, we plot $D$
of the half-filled XXZ model with impurity
density $f= 0.001$ in the attractive and the
repulsive interaction regimes as a function of impurity strength
$\nu$. In agreement with previous studies of the XXZ model in
the presence of scalar potential scatterers \cite{ramin}, 
we find that there is no enhancement in the
persistent currents for repulsive interaction 
($\Delta<0 \mbox{ i.e. } \pi/4<\eta\leq\pi/2$).  
For attractive interactions 
($\Delta>0 \mbox{ i.e. } 0\leq\eta<\pi/4$), 
we observe such a crossover  into a regime where the
currents for non-zero attraction $\eta$ are larger than the
currents of the non-interacting case at the same strength of the
impurities\cite{rarap}. Thus, although there is no
backscattering in the present case of integrable impurities, the
situation regarding enhancement of persistent currents is
qualitatively the same as for potential scatterers. 

Away from half-filling, the situation is more complicated. Here we
only remark that the degeneracy of levels observed in the clean
XXZ for $1/r$ fillings with $r$ rational \cite{rarhp} is lifted. 
Details will be published elsewhere.

%
%


In conclusion, we have shown that an interacting quantum
many-body system with impurities, that can be constructed by the
QISM, is characterised by an absence of reflection at the
impurities. As a direct consequence of the absence of back
scattering, the ground state of the interacting system remains
extended even in the presence of these special impurities. These
systems are therefore unique in that there is no 
localisation in the presence of disorder in 1D. 
The value of the effective carrier charge $-q/2$ and the 
Sutherland-Shastry relation
$D\chi=1/2\pi$ are derived for the half-filled XXX and XXZ models
and are shown to be independent of the impurity strength.
For the half-filled XXZ model, we show enhancement of the
persistent currents only for attractive interactions.

A.P.\ would like to thank Pinaki Majumdar and Ramesh Pai for 
fruitful discussions.
H.-P.E.\ and R.A.R.\ gratefully acknowledge
financial support through the European Union's "Human Capital and 
Mobility" program and the Alexander von Humboldt Foundation, respectively.

\begin{figure}
  \caption{
    Stiffness of the half-filled XXZ model with impurity
    density $f= 0.001$ for various interaction strengths $\Delta$
    as a function of impurity strength $\nu$.
    Note that increasing repulsion and impurity strength will always
    reduce $D$ whereas
    there is a crossover for all values of the attraction as we
    increase the impurity strength.
  }
\label{fig}
\end{figure}


\begin{references}
\bibitem{note1} Note that for example in the XXZ model of SSE the
        impurity phase shift $\delta$ is not only 
        a function of $k$ and $\nu$ but also of the 
        interaction strength $\Delta$.
\end{references}
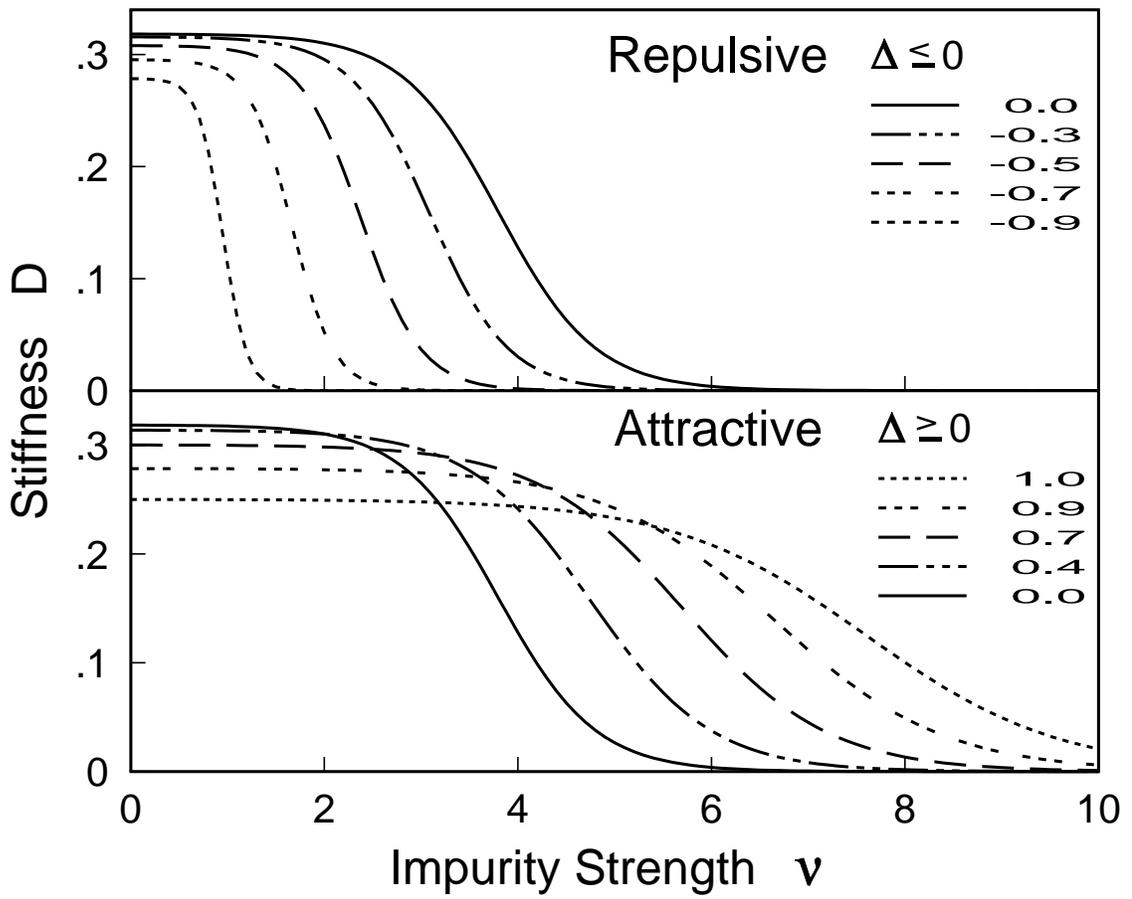
\end{document}